\documentclass[cits]{PoS}

\usepackage{amsmath}
\usepackage{multido,ifthen}
\usepackage{pstricks,pst-plot}

\title{$\eta \to 3 \pi$ and quark masses}

\ShortTitle{$\eta \to 3 \pi$ and quark masses}

\author{\speaker{Stefan Lanz}\\

		\vspace{.5ex}

		Albert Einstein Center for Fundamental Physics, Institute for Theoretical Physics,\\
		University of Bern, Sidlerstrasse 5, CH-3012 Bern, Switzerland

		\vspace{.5ex}
		
		Department of Astronomy and Theoretical Physics, Lund University,\\
		S\"olvegatan 14A, S-223 62 Lund, Sweden

		\vspace{.5ex}

    E-mail: \email{slanz@itp.unibe.ch}}

\abstract{In recent years, the decay $\eta \to 3 \pi$ has received considerable attention from experimental and theoretical side. It is of particular theoretical interest because it allows for the determination of $(m_u - m_d)$. In addition, for many years now theory has had difficulties to understand the slope of the neutral channel Dalitz plot distribution, which in contrast is very well measured. I discuss here the relation of the decay to the masses of the light quarks and review a number of theoretical and experimental works that are concerned with these questions.}

\FullConference{The 7th International Workshop on Chiral Dynamics,\\
		August 6 -10, 2012\\
		Jefferson Lab, Newport News, Virginia, USA}

\renewcommand{\logo}
	{\parbox{\textwidth}{
		\raggedleft{LU TP 13-02\\January 2013}\\[2mm]
		\includegraphics{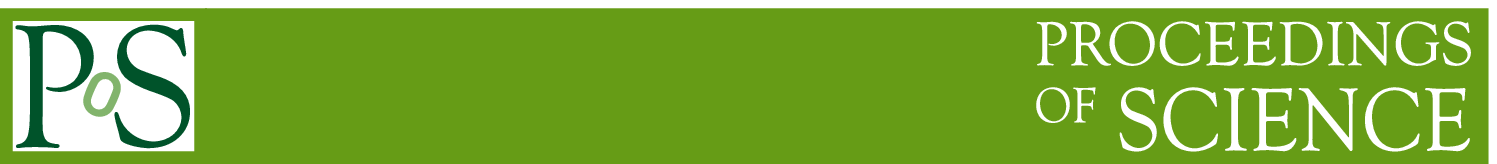}
	}
}

\FullConference{\emph{To be published in the proceedings of:}\\
		The 7th International Workshop on Chiral Dynamics,\\
		August 6 -10, 2012\\
		Jefferson Lab, Newport News, Virginia, USA}

\newcommand{\inv}[1]{\frac{1}{#1}}

\newcommand{\mpi}{m_\pi}
\newcommand{\mpip}{m_{\pi^+}}

\newcommand{\mpiz}{m_{\pi^0}}
\newcommand{\mK}{m_K}
\newcommand{\mKp}{m_{K^+}}
\newcommand{\mKz}{m_{K^0}}
\newcommand{\meta}{m_\eta}

\newcommand{\MeV}{\text{MeV}}

\newcommand{\eV}{\text{eV}}

\renewcommand{\L}{\mathcal{L}}
\renewcommand{\O}{\mathcal{O}}
\newcommand{\A}{\mathcal{A}}
\newcommand{\M}{\mathcal{M}}

\newcommand{\eolp}{\,.}
\newcommand{\eolc}{\,,}

\newlength{\dotsize}
\newcommand{\entryTextPos}{}
\newcounter{entrynum}
\newcommand{\entryNum}{0}
\newcommand{\entryFont}{\small}
\newcommand{\entry}[4]{	\addtocounter{entrynum}{-1}
								\ifthenelse{\equal{#2}{\empty}}{
									\psline{|-|}(#3,\theentrynum)(#4,\theentrynum)
								}{
									\pscircle*(#2,\theentrynum){\dotsize}
									\ifthenelse{\equal{#3}{\empty}}{}{
										\SpecialCoor
										\psline{|-|}(!#2 #3 sub \theentrynum)(!#2 #4 add \theentrynum)
										\NormalCoor
									}
								}
								\rput[l](\entryTextPos,\theentrynum){\entryFont #1}
							}

\newcommand{\dotybars}[4][5pt]{\psdot[dotsize=#1](#2,#3)
		\SpecialCoor
		\psline{|-|}(!#2 #3 #4 sub)(!#2 #3 #4 add)
		\NormalCoor
}

\newgray{fillgray}{0.8}
\definecolor{pred}{rgb}{0.7,0,0}
\definecolor{pgreen}{rgb}{0,.6,0}
\definecolor{pblue}{rgb}{0,0,.6}
\definecolor{porange}{rgb}{1.0,0.5,0.0}

\newcommand{\mylw}{0.8pt}

\begin{document}

\section{Introduction}

The masses of the light quarks, i.e., the up-, down-, and strange-quark, are not directly accessible to experimental determination due to confinement, which prevents quarks from appearing as free particles. Since these masses are much smaller than the typical scale of QCD at around 1 GeV, their contribution to hadronic quantities as, e.g., the nucleon mass is rather small. There is, however, a prominent exception to this rule, formulated in the famous Gell-Mann--Oakes--Renner relation~\cite{Gell-Mann+1968}. It states that the masses of the lightest mesons are determined by the combined effects of spontaneous and explicit chiral symmetry breaking, that is by the chiral quark condensate and the light quark masses. At leading order in the quark mass expansion and including $\eta$-$\pi^0$ mixing and first order electromagnetic corrections, one finds
\begin{align}\begin{aligned}
	\mpiz^2 &= B_0 (m_u+m_d) + \frac{2 \epsilon}{\sqrt{3}} B_0 (m_u - m_d) + \ldots \eolc
	&\mpip^2 &= B_0 (m_u+m_d) + \Delta_\textit{em}^\pi + \ldots \eolc \\
	\mKz^2 &= B_0 (m_d+m_s) + \ldots \eolc
	&\mKp^2 &= B_0 (m_u+m_s) + \Delta_\textit{em}^K + \ldots \eolc \\
	\meta^2 &= B_0 \frac{m_u+m_d + 4 m_s}{3} - \frac{2 \epsilon}{\sqrt{3}} B_0 (m_u - m_d) + \ldots \eolp
	\label{eq:GMOR}
\end{aligned}\end{align}
The parameter $\epsilon = \sqrt{3}/4 (m_d-m_u)/(m_s-\hat m) \approx 0.015$, $\hat m = (m_u+m_d)/2$, is the $\eta$-$\pi^0$ mixing angle. According to Dashen's theorem~\cite{Dashen1969}, the electromagnetic corrections to the pion and kaon mass coincide at leading order with $\Delta_{\textit{em}}^{\pi/K} \sim (35~\MeV)^2$. Because $B_0$ is not a priori known, one can only extract ratios of quark masses from Eq.~\eqref{eq:GMOR}. Assuming Dashen's theorem to be true, one finds the famous relations first derived by Weinberg using current algebra~\cite{Weinberg1977}:
\begin{align}\begin{split}
	\frac{m_d}{m_u} \approx \frac{\mKz^2 - \mKp^2 + \mpip^2}{\mKp^2 - \mKz^2 - \mpip^2 + 2 \mpiz^2} \approx 1.79 \eolc \qquad
	\frac{m_s}{m_d} \approx \frac{\mKp^2 + \mKz^2 - \mpip^2}{\mKz^2 - \mKp^2 + \mpip^2} \approx 20.2 \eolp
	\label{eq:WeinbergRatios}
\end{split}\end{align}
Access to $m_u - m_d$ is made difficult by the fact that this small quantity enters Eq.~\eqref{eq:GMOR} only quadratically and by the possibility of Dashen violating contributions to the charged kaon mass.

Lattice calculations are able to relate the quark masses to measurable meson masses, thus leading to reliable predictions for $m_s$ and $\hat m$. I will not discuss these methods further and instead refer to the many contributions in these proceedings (e.g.~\cite{Lellouch2013, Sachrajda2013, Dudek2013}) as well as to the extensive report by FLAG~\cite{Colangelo+2011}. The determination of $(m_u - m_d)$ is however difficult also in Lattice calculations, due to the reasons discussed above: it enters quadratically or is hidden behind sizable electromagnetic corrections, which are only beginning to be studied on the Lattice~\cite{Izubuchi2013, Bernard2013}.

\section{$\eta \to 3 \pi$}

The main focus of this article is the decay process $\eta \to 3 \pi$. It can appear in two variants: either the decay goes into three neutral pions, $\eta \to 3 \pi^0$, or into a pair of charged pions together with a neutral one, $\eta \to \pi^+ \pi^- \pi^0$. I will denote the amplitude by $\A_n(s,t,u)$ for the neutral and by $\A_c(s,t,u)$ for the charged channel. The Mandelstam variables are defined as $s = (p_{\pi^+} + p_{\pi^-})$, $t = (p_{\pi^0} + p_{\pi^-})$, and $u = (p_{\pi^0} + p_{\pi^+})$ in the charged  channel, where they satisfy the relation $s+t+u=\meta^2 + 2 \mpip^2 + \mpiz^2$. Due to charge conjugation symmetry, the amplitude is symmetric under $t \leftrightarrow u$. The adaptation of these definitions to the neutral channel is obvious. The neutral amplitude is even totally symmetric in $s$, $t$, and $u$.

The particular importance of the decay $\eta \to 3 \pi$ for the determination of $(m_u - m_d)$ is due to the fact that it is forbidden by isospin symmetry. The reason is that three pions can not at the same time couple to vanishing angular momentum and zero isospin. The only operator in the QCD Lagrangian that can produce such a transition is
\begin{align}
	\L_\textit{IB} = - \frac{m_u - m_d}{2} (\bar u u - \bar d d) \eolp
	\label{eq:LIB}
\end{align}
As a consequence of being generated by this $\Delta I = 1$ operator, the decay amplitude must be proportional to \mbox{$(m_u - m_d)$} and can be used to extract this quantity. The decay width can be seen as a measure for the size of  isospin breaking in QCD.

Due to the different electric charges of the up- and the down-quark, also the electromagnetic interaction is isospin violating and can contribute to the decay width of $\eta \to 3 \pi$. These contributions have been predicted to be small~\cite{Bell+1968,Sutherland1966}, which has been confirmed by explicit one-loop calculations within Chiral Perturbation Theory (ChPT)~\cite{Baur+1996,Ditsche+2009}. Let me stress, however, that Ref.~\cite{Ditsche+2009} indicates that they might not be entirely negligible. Still, $\eta \to 3 \pi$ provides a rather clean access to isospin breaking within QCD and hence to $(m_u - m_d)$.

In ChPT, quark masses are always multiplied by $B_0$. To avoid the appearance of this parameter, it is convenient to rewrite the prefactor to the amplitude in terms of quark mass ratios. Two different conventions are in use:
\begin{align} \label{eq:QRprefactor}
	\A_{\eta \to 3 \pi} \propto B_0 (m_u - m_d) = -\inv{Q^2} \frac{\mK^2(\mK^2 - \mpi^2)}{\mpi^2} + \O(\M^3)
			= -\inv{R} (\mK^2 - \mpi^2)  + \O(\M^2) \eolc
\end{align}
with
\begin{align}
	Q^2 = \frac{m_s^2 - \hat{m}^2}{m_d^2 - m_u^2} \eolc \qquad \qquad R = \frac{m_s - \hat{m}}{m_d - m_u} \eolp
\end{align}
Note that in Eq.~\eqref{eq:QRprefactor} the term containing $Q$ is accurate up to $\O(\M^3)$, while the other one is corrected already at $\O(\M^2)$. From these definitions, one finds that the two ratios are related by
\begin{align} \label{eq:QR}
	R = 2 Q^2 \left( 1 + \frac{m_s}{\hat m} \right)^{-1} \eolp
\end{align}
The ratio $m_s/\hat m$ can be determined on the lattice. The current lattice average from FLAG is $m_s/\hat m = 27.4\pm0.4$~\cite{Colangelo+2011}. Since the decay width is basically given by the phase space integral over the amplitude squared,
\begin{align}
	\Gamma_{\eta \to 3 \pi} \propto \int | \A_{\eta \to 3 \pi}(s,t,u) |^2 \propto \inv{Q^4} \propto \inv{R^2} \eolc
\end{align}
an accurate theoretical description of the decay amplitude can be used to extract either $Q$ or $R$ by comparison with an experimental value for $\Gamma_{\eta \to 3 \pi}$. Note that this also means that finding a value for $Q$ or $R$ is equivalent to finding the correct normalisation of the decay amplitude.

Of course the aforementioned procedure can be reversed: given a reliable value for $Q$ (or $R$) the decay width can be predicted. The theoretical challenge in doing this stays the same: one needs an accurate description of the decay amplitude. The obvious choice to treat the process in question is of course ChPT. But this is not entirely successful, as one sees immediately by comparing the tree-level (or current algebra)~\cite{Cronin1967,Osborn+1970}, one-loop~\cite{Gasser+1985a} and two-loop prediction~\cite{Bijnens+2007} for the decay width%
\footnote{The tree-level and one-loop values have been taken from~\cite{Gasser+1985a}, where $Q=24.15$ is used. This value follows from Dashen's theorem with the PDG meson masses from the time the article was published. No two-loop result for the decay width is quoted in~\cite{Bijnens+2007} since there, $R$ has been calculated from the experimental value for $\Gamma$. But from their results together with $Q=24.15$ and $m_s/\hat m = 27.4$, one finds $\Gamma = 298\ \eV$.}
with the current PDG value, $\Gamma_{\eta \to \pi^+ \pi^- \pi^0} = 296 \pm 16\ \eV$~\cite{PDG2012}%
:
\begin{align} \label{eq:DecayWidthChPT}
	\Gamma_{\eta \to \pi^+ \pi^- \pi^0} = 66\ \eV + 94\ \eV + 138\ \eV + \ldots = 298\ \eV + \ldots \eolp
\end{align}
Even though the result at two-loops happens to coincide almost perfectly with experiment, the chiral series does not exhibit good convergence behaviour. Since the decay width is based on the square of the amplitude, Eq.~\eqref{eq:DecayWidthChPT} is somewhat misleading: the numbers are enhanced by interference such that, e.g., the two-loop number also contains contributions that are of $\O(p^8)$ and $\O(p^{10})$. The amplitude is converging more quickly.

It has been shown independently of ChPT that the width is enhanced by large final state rescattering effects~\cite{Roiesnel+1981}. This is mirrored in the chiral series and is a motivation to treat the process with dispersive methods, which allow to sum up the contributions from final state rescattering.

Furthermore, the theoretical understanding of the energy distribution in the neutral channel is not complete. Due to symmetry reasons, the square of the amplitude can in this situation be parametrised by a single parameter $\alpha$, which has been measured by many experiments with good mutual agreement. While experiments all find this parameter to be negative, ChPT at one- and two-loop order predicts a positive value. Also other calculations have failed to reproduce this parameter satisfactorily as can be seen in the compilation of results in Fig.~\ref{fig:alpha}.

But difficulties appear also on the experimental side. There are hints of a tension between the many measurements of the neutral channel and the only available high-statistics measurement of the charged channel by KLOE. Such a statement can be made because the charged and neutral channel decay amplitudes are related by
\begin{align} \label{eq:isoRel}
	\A_n(s,t,u) = \A_c(s,t,u) + \A_c(t,u,s) + \A_c(u,s,t) \eolc
\end{align}
if only first order isospin breaking, i.e. $\Delta I = 1$, is considered. This relation allows for certain consistency checks among experiments, which will be discussed in more detail later.

To conclude this section, I want to discuss an important property of the decay amplitude, the so-called Adler zero~\cite{Adler1965}. This soft-pion theorem states that in the $SU(2)$ chiral limit, the decay amplitude has two zeros at
\begin{align}
	p_{\pi^+} \to 0 \ \Leftrightarrow \ s = u = 0 \eolc\ t = \meta^2 \eolc \quad \text{and} \quad
	p_{\pi^-} \to 0 \ \Leftrightarrow \ s = t = 0 \eolc\ u = \meta^2 \eolp
\end{align}
Moving away from the chiral limit (but keeping $\mpip = \mpiz \equiv \mpi$), the positions of the Adler zeros are shifted by a contribution of the order of $\mpi^2$ to
\begin{align}
	s = u = \frac{4}{3} \mpi^2 \eolc\ t = \meta^2 + \frac{\mpi^2}{3} \eolc \quad \text{and} \quad
	s = t = \frac{4}{3} \mpi^2 \eolc\ u = \meta^2 + \frac{\mpi^2}{3} \eolp
\end{align}
As one expects from a $SU(2)$ soft-pion theorem, the corrections to the position of the Adler zero are of order $\mpi^2$, since the symmetry forbids large $\O(m_s)$ contributions. At one-loop order, the real part has Adler zeros at $s = u = 1.35 \mpi^2$ and $s = t = 1.35 \mpi^2$, where also the imaginary part of the amplitude is small.

\section{Dalitz plot measurements}

\begin{figure}[t]
	\centerline{\includegraphics[width=.63\textwidth]{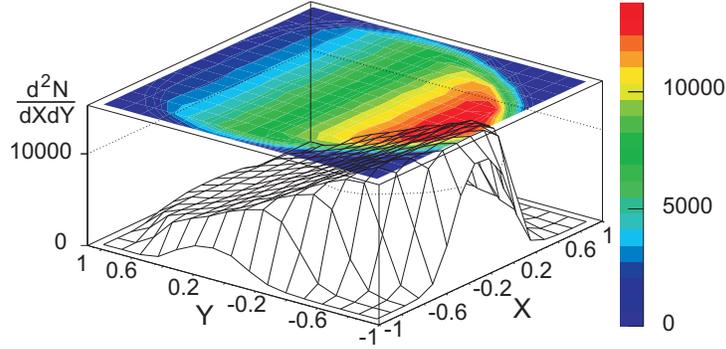}}
	\caption{The Dalitz plot distribution measured by the KLOE collaboration (figure from~\cite{Ambrosino+2008}).}
	\label{fig:DalitzKLOE}
\end{figure}

The momentum distribution of a three-particle decay is typically displayed in form of a Dalitz plot, where it is plotted against two independent kinematic variables. A common choice are the so-called Dalitz plot variables, which in the charged channel are defined by
\begin{align} \label{eq:DalitzDef}
	X = \frac{\sqrt{3}}{2 \meta Q_c} (u-t) \eolc \quad Y = \frac{3}{2 \meta Q_c} \left( (\meta - \mpiz)^2 - s \right) - 1 \eolc
\end{align}
with $Q_c = \meta - 2 \mpip - \mpiz$. In the neutral channel it is for symmetry reasons convenient to use the variable $Z = X^2 + Y^2$. The physical region of the decay process lies for both channels inside the circle with $X^2 + Y^2 = 1$ but does not entirely cover it. The point $X = Y = 0$ is referred to as the centre of the Dalitz plot. An example of a three-dimensional Dalitz plot is shown in Fig.~\ref{fig:DalitzKLOE}.

It is common to parametrise the Dalitz plot distribution as a polynomial in $X$ and $Y$. For the charged channel and up to cubic order, the Dalitz plot parametrisation reads
\begin{align} \label{eq:DalitzCharged}
	\Gamma_c(X,Y) = |\A_c(s,t,u)| \propto 1 + aY + bY^2 + cX + dX^2 + eXY + fY^3 + gX^3 + hX^2Y + lXY^2 \eolc
\end{align}
where the coefficients $a, b, \ldots$ are called Dalitz plot parameters. Charge conjugation symmetry requires terms odd in X to vanish such that $c = e = g = l = 0$.

Several experiments have measured at least some of the Dalitz plot parameters in Eq.~\eqref{eq:DalitzCharged} \cite{Gormley+1970,Layter+1973,Abele+1998a,Ambrosino+2008}, but only the measurement by KLOE has collected enough statistics to be reliable for more than one or two of them. Their result for the Dalitz plot distribution from about $1.3 \cdot 10^6$ $\eta \to \pi^+ \pi^- \pi^0$ events is shown in Fig.~\ref{fig:DalitzKLOE}. They find Dalitz plot parameters consistent with charge conjugation. Also the parameter $h$ is consistent with zero. For the others, they find
\begin{align} \label{eq:DalitzKLOE}
	a = -1.090^{+0.009}_{-0.020} \eolc \quad b = 0.124 \pm 0.012 \eolc \quad
				d = 0.057^{+0.009}_{-0.017} \eolc \quad f = 0.14 \pm 0.02 \eolp
\end{align}
This is the first time that $f$ has been measured.

The Dalitz plot measurement of the charged channel by the WASA-at-COSY collaboration, which has been announced at this conference~\cite{Coderre2013}, has since been published in a PhD thesis~\cite{Adlarson2012}. The statistics is considerably smaller than for KLOE. The new data confirms the low value for $b$, but deviates from KLOE in $d$. Also the CLAS collaboration has collected large statistics on the charged channel, but the analysis has not yet been completed~\cite{Amaryan2013}. Furthermore, a new analysis of a much larger data set from KLOE is under way that also intends to improve on certain limitations of the previous analysis (see the contribution by L.~Balkest\aa hl in Ref.~\cite{Adlarson+2012}).

The neutral channel has been measured much more often in recent years~\cite{Alde+1984,Abele+1998,Tippens+2001,Achasov+2001,Bashkanov+2007,Adolph+2009,Unverzagt+2009,Prakhov+2009,Ambrosino+2010} such that the picture is considerably clearer. Due to the symmetry in $s$, $t$, and $u$, the number of possible terms in the Dalitz plot parametrisation is much smaller in this case. Up to fourth order, it reads
\begin{align}
	\Gamma_n(X,Y) = |\A_n(s,t,u)|^2 \propto 1 + 2 \alpha Z + 6 \beta Y \left( X^2 - \dfrac{Y^2}{3} \right) + 2 \gamma Z^2 \eolp
\end{align}
So far, experiments have not reached the accuracy needed to determine $\beta$ and $\gamma$, but $\alpha$ has been measured by many experiments in good agreement, leading to the average $\alpha = -0.0315 \pm 0.0015$ \cite{PDG2012}. All the measurements entering this number are compiled in Fig.~\ref{fig:alpha}.

\section{Theoretical work}

During the last few years, $\eta \to 3 \pi$ has also received considerable attention from the theoretical side. I will in the following briefly discuss a few of the relevant works. An exhaustive discussion of the literature available on this subject is however not within the scope of this short article.

\begin{itemize}
	\item The strong contribution in the isospin limit has been calculated up to the two-loop level in ChPT~\cite{Bijnens+2007}. The corrections to the one-loop result are found to be sizable (see also Eq.~\eqref{eq:DecayWidthChPT}). Unfortunately, a large number of low-energy constants enter, some of which are not well known. The size of the $p^6$ contribution can therefore not be estimated reliably at the present stage.

	The ChPT amplitude is used to extract a value for the quark mass ratio $R = 41.3$. Using Eq.~\eqref{eq:QR} together with the FLAG average for $m_s/\hat m$, this leads to $Q=24.2$. Furthermore, the result for the neutral channel slope parameter is positive, $\alpha = 0.013 \pm 0.032$, but due to the large uncertainty also encompasses negative values.

	Based on Eq.~\eqref{eq:isoRel}, an upper limit for $\alpha$ in terms of charged channel Dalitz plot parameters is derived:
	\begin{align} \label{eq:alphaUpper}
		\alpha \leq \frac{1}{4} \left( b + d -\frac{1}{4} a^2 \right) \eolp
	\end{align}
	\looseness-1 It is this relation that can be used to check the consistency of charged and neutral channel experiments. The Dalitz plot parameters from KLOE and WASA-at-COSY both lead to an upper limit that is negative but larger than the PDG average for $\alpha$. While for KLOE the relation becomes almost an equality ($\alpha < -0.029$), the upper limit is larger for the WASA-at-COSY result ($\alpha < -0.0036$). The reason for this is mainly the larger value for $d$ in the latter case.
	
	\item The complete electromagnetic corrections up to order $p^4$ and $e^2 m_q$ have been calculated within ChPT in Ref.~\cite{Ditsche+2009}. Compared to an earlier similar calculation~\cite{Baur+1996}, the terms of order $e^2 (m_u-m_d)$ have been added. Contrary to the assumption made in the older publication, these terms are of comparable size to other $e^2 m_q$ contributions and can not simply be neglected. However, the total electromagnetic contribution is still shown to be small; of the order of a few percent in the amplitude, less than one percent in the value of $Q$.
	
	\item The decay has been analysed within the framework of non-relativistic effective field theory (NREFT) up to two loops~\cite{Schneider+2011}. The method has before been successfully applied to \mbox{$K \to 3 \pi$}~\cite{Colangelo+2006a,Bissegger+2008,Bissegger+2009} and consists basically of an expansion in small pion three-momenta around the centre of the Dalitz plot. The calculation requires two main inputs. The low-energy constants appearing in the tree-level NREFT decay amplitude are determined by matching to the one-loop amplitude from ChPT at the centre of the Dalitz plot. Final state $\pi \pi$ rescattering is included in the NREFT calculation. The additional low-energy constants that appear are fixed from two Roy equation analyses (\cite{Ananthanarayan+2001,Colangelo+2001} and~\cite{Kaminski+2008}).
	In this way, a representation of the shape of the Dalitz plot distribution in both channels is constructed. Since the overall normalisation of the amplitude can not be reliably determined within the NREFT framework, no value for the quark mass ratio $Q$ is given. A particular advantage of the calculation is that isospin-breaking effects have been included.

	It is found that rescattering effects lead to sizable corrections to the Dalitz plot parameters. One- and two-loop contributions are in general of similar size, thus emphasising the particular importance of rescattering effects in this process. Especially in the case of $\alpha$, where the already positive tree-level value is further shifted in the positive direction by the one loop correction, the two-loop contribution is large and negative. This leads to $\alpha = -0.025 \pm 0.005$, in marginal agreement with experiment. Regarding isospin breaking, it is found that the most sizable corrections to the Dalitz plot parameters are kinematic effects due to the fact that the charged and neutral pion masses are not the same.

	Using the NREFT method, it is possible to turn the upper limit in Eq.~\eqref{eq:alphaUpper} into an equality. This is achieved by expressing the required correction in terms of the Dalitz plot parameter $a$. Since information on $\eta \to 3 \pi$ enters the NREFT amplitude through the Dalitz plot parameters, no input from one-loop ChPT is needed in this case. The Dalitz plot parameters from KLOE lead to $\alpha = -0.059 \pm 0.007$ in clear disagreement with the PDG average as well as KLOE's own value~\cite{Ambrosino+2010}. As the possible source of the problem, the parameter $b$ is identified, since $b_\text{NREFT} = 0.308 > b_\text{KLOE} = 0.124$. However, the new measurement by WASA-at-COSY indicates that $d$ rather than $b$ might be responsible. Indeed, from their Dalitz plot parameters one finds $\alpha = -0.033 \pm 0.03$, which encompasses the PDG average. Clearly, more data is needed in order to finally settle this matter.
	
	\item The process has been treated with a mostly analytical dispersive approach~\cite{Kampf+2011}, where two rescattering processes are taken into account. It has to be noted that chiral power counting is strictly followed such that the number of subtraction constants coincides with the number of low-energy constants at two-loop order which is six. Also, the dispersive result can be matched exactly to the two loop-result by an appropriate choice of subtraction constants.

	The main result of this work is a dispersive representation where the subtraction constants are fitted to the charged channel data from the KLOE collaboration. Since the overall normalisation depends on $Q$, it can not be obtained from the data. Instead, the imaginary part of the dispersive amplitude along the line $t=u$ from zero up to threshold is required to be close to the two-loop result. The motivation for this procedure is that the $\O(p^6)$ corrections happen to be rather small along this line. 	The resulting amplitude reproduces the experimental Dalitz plot distribution in the physical region and leads to $Q = 23.1 \pm 0.7$ and $\alpha = -0.0044 \pm 0.004$.

	There is however a severe problem with the dispersive amplitude. As can be seen in Fig.~\ref{fig:KKNZamplitude}, it has no Adler zero, which means that it is not consistent with $SU(2)$ chiral symmetry. It is hard to justify any use of ChPT in a calculation that so blatantly violates the symmetry at its foundation.
\end{itemize}

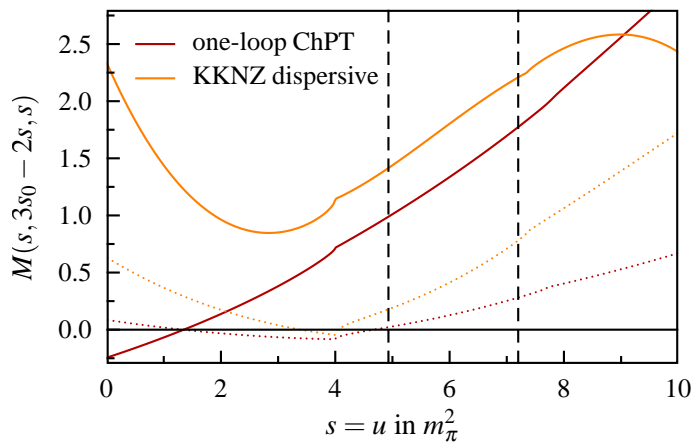
\begin{figure}[t]

	\begin{center}
	\psset{xunit=7.6mm,yunit=15.2mm}
	\begin{pspicture*}(-1.65,-0.98)(10.3,2.81)

	\psset{linewidth=\mylw,plotstyle=curve}

	% one-loop ChPT
	\psset{linecolor=pred}
	\readdata{\data}{data/ReM1loop.dat}
	\dataplot{\data}
	\readdata{\data}{data/ImM1loop.dat}
	\dataplot[linestyle=dotted,dotsep=.05]{\data}

	% KKNZ
	\psset{linecolor=porange}
	\readdata{\data}{data/ReKKNZs=u.dat}
	\dataplot{\data}
	\readdata{\data}{data/ImKKNZs=u.dat}
	\dataplot[linestyle=dotted,dotsep=.05]{\data}

	\psset{linecolor=black}

	% hide lines outside of box
	\psframe[linecolor=white,fillstyle=solid](-0.2,-0.3)(0.0,2.7)
	\psframe[linecolor=white,fillstyle=solid](10.0,-0.3)(10.5,2.7)
	\psframe[linecolor=white,fillstyle=solid](0.0,2.8)(10.5,2.9)

	% axes
	\psframe(0.0,-0.3)(10.0,2.8)
	\psline(0.0,0.0)(10.0,0.0)
	% x-axis
	\multips(2.0,-0.3)(2.0,0){4}{\psline(0,5pt)}
	\multips(1.0,-0.3)(2.0,0){5}{\psline(0,3pt)}
	\multido{\n=0+2}{6}{\uput{.2}[270](\n,-0.3){\small \n}}
	% y-axis
	\multips(0,0)(0,0.5){6}{\psline(5pt,0)}
	\multips(0,-0.25)(0,0.5){7}{\psline(3pt,0)}
	\multido{\n=0.0+0.5}{6}{\uput{0.2}[180](0,\n){\small\n}}
	% axis labels
	\uput{0.6}[270](5,-0.3){$s=u\ \text{in}\ \mpi^2$}
	\uput{.9}[180]{90}(0,1.25){$M(s,3 s_0-2 s,s)$}

	% physical region
	\psline[linestyle=dashed](7.20,-0.3)(7.20,2.8)
	\psline[linestyle=dashed](4.93,-0.3)(4.93,2.8)

	% legend
	%\psframe[linecolor=white,fillstyle=solid](2.0,2.05)(6.0,2.7)
	\psline[linecolor=pred](0.5,2.5)(1.2,2.5) \rput[l](1.5,2.5){\small one-loop ChPT}
	\psline[linecolor=porange](0.5,2.2)(1.2,2.2) \rput[l](1.5,2.2){\small KKNZ dispersive}

	\end{pspicture*}
	\end{center}
	\caption{The one-loop amplitude from ChPT together with the dispersive amplitude from Ref.~\cite{Kampf+2011} along the line $s=u$. Dashed lines represent real parts, dotted lines imaginary parts. It is clearly visible that the dispersive amplitude has no Adler zero. I thank K.~Kampf for providing the program that produced this data.}
	\label{fig:KKNZamplitude}
\end{figure}

\section{Our dispersive analysis}

I want to conclude the discussion of the theoretical literature on $\eta \to 3 \pi$ by a somewhat more detailed description of an ongoing numerical dispersive calculation in collaboration with G.~Colangelo, H.~Leutwyler, and E.~Passemar. Intermediate status reports of this work have been presented at conferences before~\cite{Adlarson+2012,Colangelo+2009,Colangelo+2011a} and have been published in form of a PhD thesis~\cite{Lanz2011}. A more detailed discussion of the formalism can be found in these references and  I will focus here on newer developments that are not described there.

The method we apply has been described in detail in Refs.~\cite{Anisovich+1996,Walker1998}. Because the calculation is numerical, it is possible to include an arbitrary number of rescattering processes. Also, chiral power counting is not followed: we always use the best available input, e.g., we do not expand the $\pi \pi$ scattering input to the appropriate order in each iteration. The method involves two main steps that can be treated entirely independently: A dispersive representation must be derived and solved numerically. The result is an amplitude that is a function of a number of unknown subtraction constants. The second step is then to determine these constants in a good way.

The dispersion relations are based on the decomposition of the normalised decay amplitude as~\cite{Stern+1993,Anisovich+1996}
\begin{align} \label{eq:Mdecomp}
	\M(s, t, u) = M_0(s) + (s-u) M_1(t) + (s-t) M_1(u) + M_2(t) + M_2(u) - \frac{2}{3} M_2(s) \eolc
\end{align}
where the subscript stands for the isospin of the scattered pion pair. This decomposition has the advantage that one deals with functions of a single variable only. Using analyticity and unitarity, one arrives at a set of dispersion relations for the three functions $M_I(s)$ that are all of the form
\begin{align} \label{eq:dispRel}
	M_I(s) = \Omega_I(s) \left\{ P_I(s) + \frac{s^{n_I}}{\pi} \int _{4 m_\pi^2}^{\infty} \frac{ds^\prime}{s^{\prime n_I}}
		\frac{\sin \delta_I(s^\prime) \hat{M}_I(s^\prime)}{|\Omega_I(s^\prime)| (s^\prime - s -i \epsilon)}
		\right\} \eolc
\end{align}
where $\Omega_I(s)$ are the so-called Omn\`es functions~\cite{Omnes1958} given by
\begin{align}
	\Omega_I(s) = \exp \left\{ \frac{s}{\pi} \int\limits_{4 \mpi^2}^\infty
							\!\! ds'\, \frac{\delta_I(s')}{s'(s'-s - i \epsilon)} \right\} \eolp
\end{align}
The functions $\hat{M}_I(s)$ are angular averages over \emph{all} the $M_I(s)$, such that the dispersion relations are recursive and coupled with each other. Two kinds of inputs are needed. On the one hand, one needs to know the $\pi\pi$ scattering phase shifts $\delta_I(s)$, which we take from Ref.~\cite{Colangelo+2001}. On the other hand, the polynomials $P_I(s)$ contain the  subtraction constants, which must be determined with information from outside the dispersive machinery. It is on this issue that progress has taken place recently.

The dispersive representations in Eq.~\eqref{eq:dispRel} can be Taylor expanded:
\begin{align}
	M_I(s) = a_I + b_I s + c_I s^2 + d_I s^3 + \ldots \eolp
\end{align}
The $M_I(s)$ are uniquely determined once all the subtraction constants are fixed. They have unique Taylor expansions, such that one can define a unique relation among a set of subtraction constants and an appropriate equally sized set of Taylor coefficients. A solution of the dispersion relations can therefore be specified by giving values for the subtraction constants \emph{or} for the Taylor coefficients. The advantage of working with the latter is that they pick up imaginary parts only at $\O(p^6)$, such that they can safely be approximated as real. This then automatically leads to subtraction constants, where the imaginary part is non-zero, but suppressed compared to the real part.

The splitting of the amplitude into the $M_I(s)$ is not unique because of the relation $s+t+u=\meta^2 + 2 \mpip^2 + \mpiz^2$: one can add polynomials to the $M_I(s)$ in such a way that $M(s,t,u)$ is not changed. This gauge freedom allows to choose some Taylor coefficients arbitrarily and the number of free parameters is therefore smaller than the number of Taylor coefficients (or subtraction constants) that are used.

We have tried to work with different numbers of subtraction constants, but comparison with data has shown that satisfactory agreement is only achieved, if eleven constants are used. The corresponding Taylor expansions read
\begin{align}\begin{aligned}
	M_0(s) &= {\color{pred}a_0} + {\color{pred}b_0} s + {\color{pblue}c_0} s^2 + {\color{porange}d_0} s^3 + \ldots \eolc\\
	M_1(s) &= {\color{pred}a_1} + {\color{pblue}b_1} s + {\color{porange}c_1} s^2 + \ldots \eolc \\
	M_2(s) &= {\color{pred}a_2} + {\color{pblue}b_2} s + {\color{pblue}c_2} s^2 + {\color{pgreen}d_2} s^3 + \ldots \eolp
\end{aligned}\end{align}
The reason for ending the series one order lower for $M_1(s)$ is that this function is multiplied by another power of momenta in Eq.~\eqref{eq:Mdecomp}, such that all three functions lead to contributions of $\O(s^3)$ in the total amplitude.

We determined the subtraction constants with two different methods. The first one relies entirely on one-loop ChPT as input and makes no use of data at all:
\begin{itemize}
	\item Since one-loop ChPT can not be trusted for terms beyond $\O(s^2)$, the coefficients ${\color{porange}d_0}$, ${\color{porange}c_1}$, and ${\color{pgreen}d_2}$ are set to zero. The dispersion relations thus contain eight subtraction constants.
	\item The parameters ${\color{pred}a_0}$, ${\color{pred}b_0}$, ${\color{pred}a_1}$, and ${\color{pred}a_2}$ are set to their tree-level value using the gauge freedom. This means that there are actually only four free parameters in the dispersion relations.
	\item Finally, the remaining parameters ${\color{pblue}c_0}$, ${\color{pblue}b_1}$, ${\color{pblue}b_2}$, and ${\color{pblue}c_2}$ are set to their one-loop value.
\end{itemize}
In this way, the dispersive solution is entirely fixed. It will later be referred to as ``dispersive, one loop''. I stress again that for this solution, only ChPT at low energy is used in order to fix four subtraction constants. This low-energy information is then extrapolated to the physical region by means of the dispersion relations.

The second method uses the full set of eleven subtraction constants:
\begin{itemize}
	\item The parameters ${\color{pred}a_0}$, ${\color{pred}b_0}$, ${\color{pred}a_1}$, ${\color{pred}a_2}$, ${\color{pblue}c_0}$, ${\color{pblue}b_1}$, ${\color{pblue}b_2}$, and ${\color{pblue}c_2}$ are determined exactly as before.
	\item The presence of more parameters also enlarges the gauge freedom since polynomials of higher order can now be added to the $M_I(s)$. The parameter ${\color{pgreen}d_2}$ is chosen such that $\delta_2$, which is the coefficient to $s^3$ in $P_2(s)$, is zero. The  number of free parameters is thus six in this case.
	\item Finally, the remaining two parameters ${\color{porange}d_0}$ and ${\color{porange}c_1}$ are determined by fitting the square of the amplitude in the physical region to the charged channel data from the KLOE collaboration.
\end{itemize}
This leads to another solution of the dispersion relations which will be referred to as ``dispersive, fit to KLOE''. We have also fitted other available data sets, but for simplicity, only one of these fits is presented here as an example.

\begin{figure}[t]

% Dalitz plot distribution for the charged channel along the line X = 0 as a function of Y
\psset{xunit=2.85cm,yunit=1.25cm}
\begin{pspicture*}(-1.45,-0.6)(1.13,3.05)

\psset{linewidth=\mylw,plotstyle=curve}

% one loop
\readdata{\data}{data/dalitzYc1loop.dat}
\dataplot[linecolor=pred]{\data}

% KLOE
\psset{linecolor=black}
\readdata{\data}{data/dalitzYcKLOECentral.dat}
\dataplot{\data}
\readdata{\data}{data/dalitzYcKLOEUpper.dat}
\dataplot[linestyle=dotted,dotsep=.05]{\data}
\readdata{\data}{data/dalitzYcKLOELower.dat}
\dataplot[linestyle=dotted,dotsep=.05]{\data}

% dispersive with matching
\psset{linecolor=pgreen}
\readdata{\data}{data/dalitzYcMatchCentral.dat}
\dataplot{\data}

% dispersive with fit
\psset{linecolor=pblue}
\readdata{\data}{data/dalitzYcKfitCentral.dat}
\dataplot{\data}

\psset{linecolor=black}

% axes
\psframe(-1.002,0.1)(1.002,2.98)
% x-axis
\multips(-.5,0.1)(0.5,0){3}{\psline(0,5pt)}
\multips(-0.75,0.1)(0.5,0){4}{\psline(0,3pt)}
\multido{\n=-1.0+0.5}{5}{\uput{.2}[270](\n,0.15){\small \n}}
% y-axis
\multips(-1,0.5)(0,0.5){5}{\psline(5pt,0)}
\multips(-1,0.25)(0,0.5){6}{\psline(3pt,0)}
\multido{\n=0.5+0.5}{5}{\uput{0.18}[180](-1,\n){\small \n}}
% axis labels
\uput{0.55}[270](0,0.1){$Y$}
\uput{0.9}[180]{90}(-1,1.45){$\Gamma(0,Y)$}

% physical region
%\psline[linestyle=dashed](-1,0.1)(-1,2.98)
\psline[linestyle=dashed](0.895,0.1)(0.895,2.98)

% legend
\psframe[fillstyle=solid](-0.2,1.7)(1.13,3.05)
\psline[linecolor=pred](-0.15,2.8)(0.0,2.8) \rput[l](0.05,2.8){\footnotesize one-loop ChPT}
\psline[linecolor=black](-0.15,2.5)(0.0,2.5) \rput[l](0.05,2.5){\footnotesize KLOE}
\psline[linecolor=pgreen](-0.15,2.2)(0.0,2.2) \rput[l](0.05,2.2){\footnotesize dispersive, one loop}
\psline[linecolor=pblue](-0.15,1.9)(0.0,1.9) \rput[l](0.05,1.9){\footnotesize dispersive, fit to KLOE}

%preliminary
\rput[l](-0.85,2.7){\color{pred} \small preliminary}

\end{pspicture*}
\hfill
\psset{xunit=5.5cm,yunit=34.0cm}
\begin{pspicture*}(-0.25,0.897)(1.06,1.031)

\psset{linewidth=\mylw,plotstyle=curve}

% one loop
\readdata{\data}{data/dalitzZn1loop.dat}
\dataplot[linecolor=pred]{\data}

% dispersive with matching
\psset{linecolor=pgreen}
\readdata{\data}{data/dalitzZnMatchCentral.dat}
\dataplot{\data}

% dispersive with fit
\psset{linecolor=pblue}
\readdata{\data}{data/dalitzZnKfitCentral.dat}
\dataplot{\data}

\psset{linecolor=black,linewidth=\mylw}
% MAMI-C data
\newcommand{\psize}{3.5pt}
\dotybars[\psize]{0.025}{0.9969}{0.002378}
\dotybars[\psize]{0.075}{0.9965}{0.002388}
\dotybars[\psize]{0.125}{0.993534}{0.00238908}
\dotybars[\psize]{0.175}{0.995036}{0.00239911}
\dotybars[\psize]{0.225}{0.98515}{0.00239309}
\dotybars[\psize]{0.275}{0.987081}{0.00240312}
\dotybars[\psize]{0.325}{0.98652}{0.00240914}
\dotybars[\psize]{0.375}{0.983344}{0.00241014}
\dotybars[\psize]{0.425}{0.978833}{0.00241415}
\dotybars[\psize]{0.475}{0.973555}{0.00241516}
\dotybars[\psize]{0.525}{0.974874}{0.00242519}
\dotybars[\psize]{0.575}{0.967591}{0.00241917}
\dotybars[\psize]{0.625}{0.964004}{0.00241917}
\dotybars[\psize]{0.675}{0.967942}{0.00243522}
\dotybars[\psize]{0.725}{0.955205}{0.00242117}
\dotybars[\psize]{0.775}{0.954161}{0.002701}
\dotybars[\psize]{0.825}{0.948984}{0.00316538}
\dotybars[\psize]{0.875}{0.947182}{0.00363377}
\dotybars[\psize]{0.925}{0.935755}{0.0042857}
\dotybars[\psize]{0.975}{0.938297}{0.00604291}

% axes
\psframe(0.0,0.921)(1.002,1.028)
% x-axis
\multips(0.25,0.921)(0.25,0){3}{\psline(0,5pt)}
\multips(0.125,0.921)(0.25,0){4}{\psline(0,3pt)}
\multido{\n=0.00+0.25}{5}{\uput{.15}[270](\n,0.921){\small \n}}
% y-axis
\multips(0,0.94)(0,0.04){4}{\psline(5pt,0)}
\multips(0,0.96)(0,0.04){4}{\psline(3pt,0)}
\multido{\n=0.94+0.04}{4}{\uput{0.15}[180](0,\n){\small \n}}
% axis labels
\uput{0.55}[270](0.5,0.921){$Z$}
\uput{0.9}[180]{90}(0,0.978){$\Gamma(Z)$}

% legend
\psline[linecolor=pred](0.05,0.961)(0.12,0.961) \rput[l](0.14,0.960){\footnotesize one-loop ChPT}
\dotybars[\psize]{0.085}{0.951}{0.003} \rput[l](0.14,0.951){\footnotesize MAMI-C}
\psline[linecolor=pgreen](0.05,0.941)(0.12,0.941) \rput[l](0.14,0.941){\footnotesize dispersive, one loop}
\psline[linecolor=pblue](0.05,0.931)(0.12,0.931) \rput[l](0.14,0.931){\footnotesize dispersive, fit to KLOE}

%preliminary
\rput[l](0.05,1.019){\color{pred} \small preliminary}

\end{pspicture*}

\caption{Results from the numerical dispersive analysis. \emph{Left panel:} The square of the amplitude in the charged channel along the line $X=0$ in comparison with the KLOE curve. The dashed line marks the end of the physical region. \emph{Right panel:} The square of the amplitude in the neutral channel integrated along circles of constant $Z$ in comparison with data points from MAMI-C. Note that none of the theoretical curves involves experimental information on the neutral channel.}
\label{fig:dispDalitz}
\end{figure}
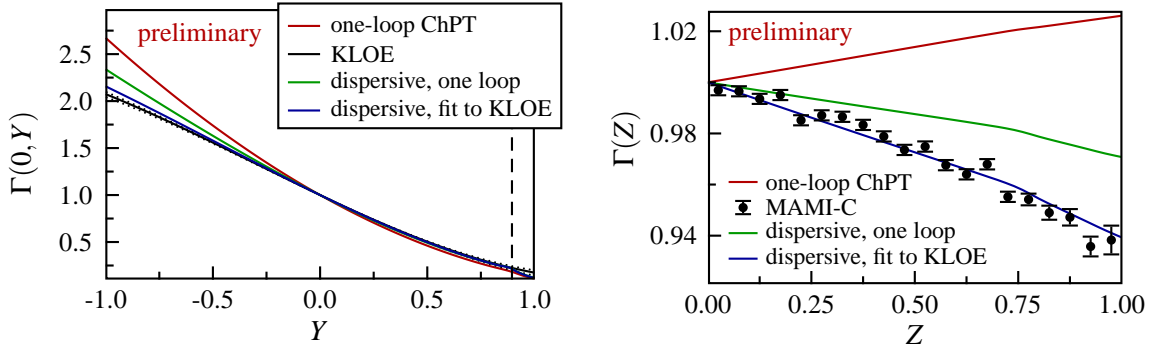

The Dalitz plot distributions from both solutions are depicted in Fig.~\ref{fig:dispDalitz} together with the corresponding one-loop results. The left panel shows the $Y$ distribution along the line with $X=0$, which is equivalent to the $s$ distribution for $t=u$. Note that small $s$ correspond to large $Y$ due to the minus sign in the definition of $Y$ in Eq.~\eqref{eq:DalitzDef}. Clearly, ChPT is successful at low energy, but fails towards the upper end of the physical region. It is noteworthy that the extrapolation of the low-energy ChPT amplitude through the dispersion relation already leads to a considerably better agreement with experiment, which is then further improved by the fit. It seems not, at first sight, remarkable that the fit does agree with the data it is fitted to. However, the fit must at the same time be consistent with the chiral constraints at low energy and it is important to show that this can actually be achieved.

Through Eq.~\eqref{eq:isoRel}, the neutral channel amplitude can be calculated from the charged channel. The right panel of Fig.~\ref{fig:dispDalitz} shows the $Z$ distributions that one finds in this way from the three charged channel amplitudes. Accordingly, the blue curve involves no experimental information on the neutral channel. But the influence of the charged channel data is exactly what is needed to bring the shape of the $Z$ distribution into agreement with the neutral channel data. To visualise this, the figure contains data points from the MAMI-C collaboration~\cite{Prakhov+2009} as an example.

But not all problems are solved yet. The  dispersive amplitude is evaluated in the isospin limit, while the data are collected in the real world with two different pion masses and electromagnetic effects. These effects are expected to be small but if they are entirely neglected, we predict a neutral channel Dalitz plot distribution that is not in agreement with the data. We expect the largest isospin correction to be kinematic effects due to the pion masses. Indeed, taking these into account by shifting the corresponding singularities to their physical position through a slight deformation of the phase space, we find that a fit to charged channel data leads to good agreement with neutral channel data as well.

While isospin corrections to the shape of the Dalitz plot distribution are successfully treated in this way, the decay rate suffers from the procedure. This can be seen in the fact that we find a branching ratio $r = \Gamma_{\eta \to 3 \pi^0}/\Gamma_{\eta \to \pi^+ \pi^- \pi^0}$ that is not in agreement with experiment. But from the estimates for electromagnetic effects in Ref.~\cite{Ditsche+2009}, we expect that the one-loop result from ChPT including isospin breaking can be used to remedy the situation.

\section{Comparison of results for $\alpha$ and $Q$}

To conclude, I have compiled various experimental and theoretical results for $\alpha$ and $Q$ in Figs.~\ref{fig:alpha} and~\ref{fig:Q}. In particular, all the results from works that I have mentioned above are listed.

Figure~\ref{fig:alpha} clearly shows the failure of ChPT at one- and two-loop order to reproduce even the sign of $\alpha$. On the other hand, all four dispersive results do reproduce the sign, and our fit to KLOE is even in agreement with the PDG average. Also the NREFT calculation leads to a value that is compatible with experiment. The current PDG average, which is marked by the grey band, includes all the experimental values that are given in the figure.
%
% commands for names of matching and fit to KLOE, such that these can be easily changed
\newcommand{\matchName}{dispersive, one loop}
\newcommand{\fitName}{dispersive, fit to KLOE}
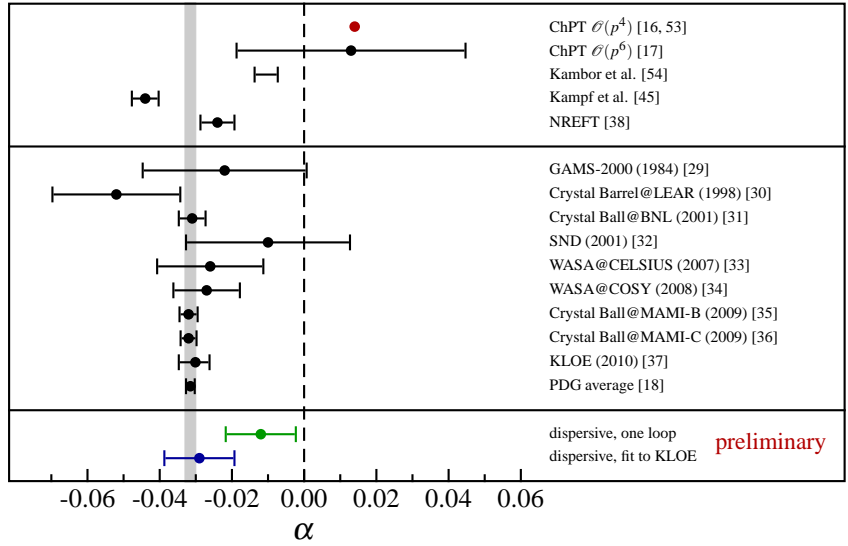
\begin{figure}[t]
\begin{center}
	\psset{xunit=48cm, yunit=.319cm}
	\begin{pspicture}(-0.082,0)(0.15,-21.7)

		% some definitions
		\setcounter{entrynum}{0}
		\renewcommand{\entryNum}{20}
		\renewcommand{\entryTextPos}{0.068}
		\setlength{\dotsize}{2.0pt}
		\renewcommand{\entryFont}{\tiny}

		\psline[linestyle=dashed](0,0)(0,-\entryNum)

		% PDG average
		\psframe[fillstyle=solid,fillcolor=fillgray,linecolor=fillgray,linewidth=0](-0.0300,-\entryNum)(-0.0330,0)

		% frame and coordinate system with label
		\psframe(-0.082,0)(0.15,-\entryNum)
		\multips(-0.06,-\entryNum)(0.02,0){7}{\psline(0,5pt)}
		\multips(-0.07,-\entryNum)(0.02,0){7}{\psline(0,3pt)}
		\multido{\n=-0.06+0.02}{7}{\uput{.1}[270](\n,-\entryNum){\small \n}}
		\uput{.55}[270](0,-\entryNum){\large $\alpha$}

		% theoretical results
		\psset{linecolor=pred}
		\entry{ChPT $\O(p^4)$ \cite{Gasser+1985a,Bijnens+2002}}{0.014}{}{}
		\psset{linecolor=black}
		\entry{ChPT $\O(p^6)$ \cite{Bijnens+2007}}{0.013}{0.032}{0.032}
		\entry{Kambor et al. \cite{Kambor+1996}}{}{-0.014}{-0.007}
		\entry{Kampf et al. \cite{Kampf+2011}}{-0.044}{0.004}{0.004}
		\entry{NREFT \cite{Schneider+2011}}{-0.024}{0.005}{0.005}

		\addtocounter{entrynum}{-1}
		\psline(-0.082,\theentrynum)(0.15,\theentrynum)

		% experimental results
		\entry{GAMS-2000 (1984) \cite{Alde+1984}}{-0.022}{0.023}{0.023}
		\entry{Crystal Barrel@LEAR (1998) \cite{Abele+1998}}{-0.052}{0.018}{0.018}
		\entry{Crystal Ball@BNL (2001) \cite{Tippens+2001}}{-0.031}{0.004}{0.004}
		\entry{SND (2001) \cite{Achasov+2001}}{-0.010}{0.023}{0.023}
		\entry{WASA@CELSIUS (2007) \cite{Bashkanov+2007}}{-0.026}{0.015}{0.015}
		\entry{WASA@COSY (2008) \cite{Adolph+2009}}{-0.027}{0.0095}{0.0095}
		\entry{Crystal Ball@MAMI-B (2009) \cite{Unverzagt+2009}}{-0.032}{0.0028}{0.0028}
		\entry{Crystal Ball@MAMI-C (2009) \cite{Prakhov+2009}}{-0.032}{0.0025}{0.0025}
		\entry{KLOE (2010) \cite{Ambrosino+2010}}{-0.0301}{0.0049}{0.0042}
		\entry{PDG average \cite{PDG2012}}{-0.0315}{0.0015}{0.0015}

		\addtocounter{entrynum}{-1}
		\psline(-0.082,\theentrynum)(0.15,\theentrynum)

		% our results
		\psset{linecolor=pgreen}
		\entry{\matchName}{-0.012}{0.01}{0.01}
		\psset{linecolor=pblue}
		\entry{\fitName}{-0.029}{0.01}{0.01}

	%preliminary
	\rput[l](0.114,-18.4){\footnotesize \color{pred} preliminary}

	\end{pspicture}
	\caption{Comparison of various theoretical and experimental results for the slope parameter $\alpha$.}
	\label{fig:alpha}
\end{center}
\end{figure}

The values for $Q$ that are shown in Fig.~\ref{fig:Q} cover the range from about 21 up to $Q_D = 24.3$, which follows from Dashen's theorem. Our preliminary value from the fit to KLOE lies around 22. I have conservatively assumed an error of one unit, but expect the final error to be smaller.

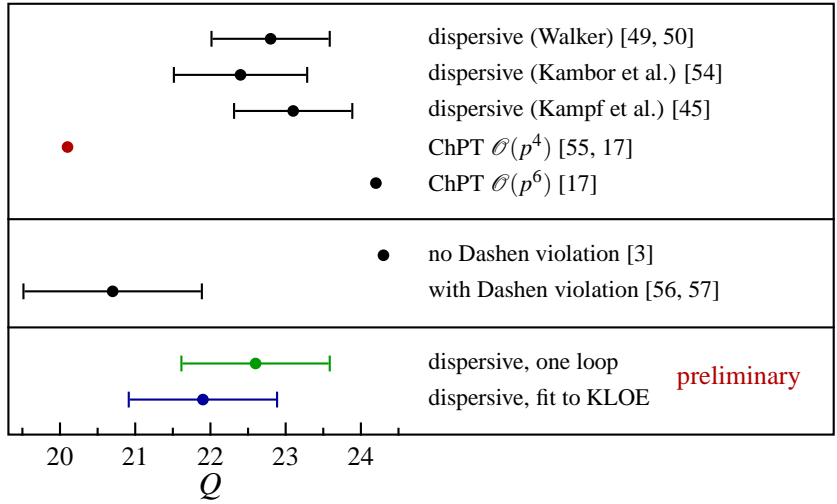
\begin{figure}[t]

\begin{center}
	\psset{xunit=1.0cm, yunit=.48cm}
	\begin{pspicture}(19.25,0)(30.5,-13.2)

		% some definitions
		\setcounter{entrynum}{0}
		\renewcommand{\entryNum}{12}
		\setlength{\dotsize}{2.2pt}
		\renewcommand{\entryTextPos}{24.9}
		\renewcommand{\entryFont}{\footnotesize}

		% frame and coordinate system with label
		\psframe(19.3,0)(30.3,-\entryNum)
		\multips(20.0,-\entryNum)(1,0){5}{\psline(0,5pt)}
		\multips(19.5,-\entryNum)(1,0){6}{\psline(0,3pt)}
		\multido{\n=20+1}{5}{\uput{.15}[270](\n,-\entryNum){\small \n}}
		\uput{.5}[270](22,-\entryNum){\large $Q$}

		% results for Q
		\entry{dispersive (Walker) \cite{Anisovich+1996,Walker1998}}{22.8}{0.8}{0.8}
		\entry{dispersive (Kambor et al.) \cite{Kambor+1996}}{22.4}{0.9}{0.9}
		\entry{dispersive (Kampf et al.) \cite{Kampf+2011}}{23.1}{0.8}{0.8}
		\psset{linecolor=pred}
		\entry{ChPT $\O(p^4)$ \cite{Gasser+1985, Bijnens+2007}}{20.1}{}{}
		\psset{linecolor=black}
		\entry{ChPT $\O(p^6)$ \cite{Bijnens+2007}}{24.2}{}{}
		\addtocounter{entrynum}{-1}
		\psline(19.3,\theentrynum)(30.3,\theentrynum)
		\entry{no Dashen violation \cite{Weinberg1977}}{24.3}{}{}
		\entry{with Dashen violation \cite{Ananthanarayan+2004,Kastner+2008}}{20.7}{1.2}{1.2}
		\addtocounter{entrynum}{-1}
		\psline(19.3,\theentrynum)(30.3,\theentrynum)
		\psset{linecolor=pgreen}
		\entry{\matchName}{22.6}{1.0}{1.0}
		\psset{linecolor=pblue}
		\entry{\fitName}{21.9}{1.0}{1.0}

	%preliminary
	\rput[l](28.2,-10.4){\small \color{pred} preliminary}
	\end{pspicture}

	\caption{Comparison of various theoretical and experimental results for the quark mass ratio $Q$.}
	\label{fig:Q}
\end{center}
\end{figure}

\section{Conclusion \& Outlook}

The process $\eta \to 3 \pi$ is a unique source of information about isospin breaking in QCD and thus for estimating $m_u-m_d$. A wealth of theoretical and experimental work has been dedicated to this decay and the question of quark mass extraction. With dispersive techniques, the sizable final state rescattering effects can be treated properly and a significant improvement of the one-loop result from ChPT has been achieved in this way. In particular, the neutral channel slope parameter can be understood based on charged channel data. Consequently, the dispersive analysis does not produce a clear sign of a tension among experiments, even though that possibility can not yet be entirely excluded. Before a final statement can be made, a more careful treatment of isospin breaking effects must be implemented.

\acknowledgments

I thank the organisers for an enjoyable and stimulating workshop. I am particularly grateful for the possibility to present the topics that are discussed in this article. G.~Colangelo, H.~Leutwyler, and E.~Passemar I thank for the fruitful collaboration, J.~Bijnens, K.~Kampf, B.~Kubis, S.~Schneider, and P.~Stoffer for useful and interesting discussions. This work is supported by a scholarship of the Swiss National Science Foundation SNF.

%\bibliographystyle{JHEPmod}
%\bibliography{../../Bibliographie/ReferencesPhysics.bib}

\begin{thebibliography}{10}

\bibitem{Gell-Mann+1968}
M.~Gell-Mann, R.~J. Oakes, and B.~Renner, {\it Behavior of current divergences
  under {$SU(3) \times SU(3)$}},  {\em Phys. Rev.} {\bf 175} (1968) 2195--2199.

\bibitem{Dashen1969}
R.~F. Dashen, {\it Chiral {SU}(3)$\times${SU}(3) as a symmetry of the strong
  interactions},  {\em Phys. Rev.} {\bf 183} (1969) 1245--1260.

\bibitem{Weinberg1977}
S.~Weinberg, {\it The problem of mass},  {\em Trans. New York Acad. Sci.} {\bf
  38} (1977) 185--201.

\bibitem{Lellouch2013}
L.~Lellouch, {\it Meson chiral perturbation theory meets lattice {QCD}},  {\em
  PoS} {\bf CD12} (2013) 08.

\bibitem{Sachrajda2013}
C.~Sachrajda, {\it Non-leptonic and rare kaon decays in lattice {QCD}},  {\em
  PoS} {\bf CD12} (2013) 009.

\bibitem{Dudek2013}
J.~Dudek, {\it Meson spectra from lattice {QCD}},  {\em PoS} {\bf CD12} (2013)
  019.

\bibitem{Colangelo+2011}
G.~Colangelo {et.~al.}, {\it Review of lattice results concerning low-energy
  particle physics},  {\em Eur.Phys.J.} {\bf C71} (2011) 1695,
  [\href{http://xxx.lanl.gov/abs/1011.4408}{{\tt arXiv:1011.4408}}].

\bibitem{Izubuchi2013}
T.~Izubuchi, {\it Isospin breaking studies from lattice {QCD} + {QED}},  {\em
  PoS} {\bf CD12} (2013) 026.

\bibitem{Bernard2013}
C.~Bernard, {\it Electromagnetic contributions to pseudoscalar masses},  {\em
  PoS} {\bf CD12} (2013) 030.

\bibitem{Bell+1968}
J.~S. Bell and D.~G. Sutherland, {\it Current algebra and $\eta \to 3 \pi$},
  {\em Nucl. Phys.} {\bf B4} (1968) 315--325.

\bibitem{Sutherland1966}
D.~G. Sutherland, {\it Current algebra and the decay $\eta \to 3 \pi$},  {\em
  Phys. Lett.} {\bf 23} (1966) 384.

\bibitem{Baur+1996}
R.~Baur, J.~Kambor, and D.~Wyler, {\it Electromagnetic corrections to the
  decays $\eta \to 3\pi$},  {\em Nucl. Phys.} {\bf B460} (1996) 127--142,
  [\href{http://xxx.lanl.gov/abs/hep-ph/9510396}{{\tt hep-ph/9510396}}].

\bibitem{Ditsche+2009}
C.~Ditsche, B.~Kubis, and U.-G. Mei\ss{}ner, {\it Electromagnetic corrections
  in $\eta \to 3 \pi$ decays},  {\em Eur. Phys. J.} {\bf C60} (2009) 83--105,
  [\href{http://xxx.lanl.gov/abs/0812.0344}{{\tt arXiv:0812.0344}}].

\bibitem{Cronin1967}
J.~A. Cronin, {\it Phenomenological model of strong and weak interactions in
  chiral ${U}(3) \times {U}(3)$},  {\em Phys. Rev.} {\bf 161} (1967)
  1483--1494.

\bibitem{Osborn+1970}
H.~Osborn and D.~J. Wallace, {\it $\eta$-${X}$ mixing, $\eta \to 3 \pi$ and
  chiral {L}agrangians},  {\em Nucl. Phys.} {\bf B20} (1970) 23--44.

\bibitem{Gasser+1985a}
J.~Gasser and H.~Leutwyler, {\it $\eta \to 3 \pi$ to one loop},  {\em Nucl.
  Phys.} {\bf B250} (1985) 539.

\bibitem{Bijnens+2007}
J.~Bijnens and K.~Ghorbani, {\it $\eta \to 3 \pi$ at two loops in chiral
  perturbation theory},  {\em JHEP} {\bf 0711} (2007) 030,
  [\href{http://xxx.lanl.gov/abs/0709.0230}{{\tt arXiv:0709.0230}}].

\bibitem{PDG2012}
{\bf Particle Data Group} Collaboration, J.~Beringer {et.~al.}, {\it Review of
  particle physics},  {\em Phys. Rev.} {\bf D86} (2012) 010001.

\bibitem{Roiesnel+1981}
C.~Roiesnel and T.~N. Truong, {\it Resolution of the $\eta \to 3 \pi$ problem},
   {\em Nucl. Phys.} {\bf B187} (1981) 293--300.

\bibitem{Adler1965}
S.~L. Adler, {\it Consistency conditions on the strong interactions implied by
  a partially conserved axial-vector current},  {\em Phys. Rev.} {\bf 137}
  (1965) B1022--B1033.

\bibitem{Ambrosino+2008}
{\bf KLOE} Collaboration, F.~Ambrosino {et.~al.}, {\it Determination of
  $\eta\to\pi^+\pi^-\pi^0$ {D}alitz plot slopes and asymmetries with the {KLOE}
  detector},  {\em JHEP} {\bf 05} (2008) 006,
  [\href{http://xxx.lanl.gov/abs/0801.2642}{{\tt arXiv:0801.2642}}].

\bibitem{Gormley+1970}
M.~Gormley {et.~al.}, {\it Experimental determination of the {D}alitz-plot
  distribution of the decays $\eta \to \pi^+ \pi^- \pi^0$ and $\eta \to \pi^+
  \pi^- \gamma$, and the branching ratio $\eta \to \pi^+ \pi^- \gamma / \eta
  \to \pi^+ \pi^- \pi^0$},  {\em Phys. Rev. D} {\bf 2} (1970) 501--505.

\bibitem{Layter+1973}
J.~G. Layter {et.~al.}, {\it Study of {D}alitz-plot distributions of the decays
  $\eta \to \pi^+ \pi^- \pi^0$ and $\eta \to \pi^+ \pi^- \gamma$},  {\em Phys.
  Rev. D} {\bf 7} (May, 1973) 2565--2568.

\bibitem{Abele+1998a}
{\bf Crystal Barrel} Collaboration, A.~Abele {et.~al.}, {\it Momentum
  dependence of the decay $\eta \to \pi^+ \pi^- \pi^0$},  {\em Phys. Lett.}
  {\bf B417} (1998) 197--201.

\bibitem{Coderre2013}
D.~Coderre, {\it Tests of symmetries with eta decays at {WASA-at-COSY}},  {\em
  PoS} {\bf CD12} (2013) 063.

\bibitem{Adlarson2012}
P.~Adlarson, {\it Studies of the decay $\eta \to \pi^+ \pi^- \pi^0$ with
  {WASA}-at-{COSY}}.
\newblock PhD thesis, Uppsala University, 2012.

\bibitem{Amaryan2013}
M.~Amaryan, {\it Photoproduction and decay of light mesons in {CLAS}},  {\em
  PoS} {\bf CD12} (2013) 061.

\bibitem{Adlarson+2012}
P.~Adlarson {et.~al.}, {\it
  Proceedings of the {S}econd {I}nternational {P}rime{N}et {W}orkshop},
  \href{http://xxx.lanl.gov/abs/1204.5509}{{\tt arXiv:1204.5509}}.

\bibitem{Alde+1984}
{\bf Serpukhov-Brussels-Annecy(LAPP)} Collaboration, D.~Alde {et.~al.}, {\it
  Neutral decays of the $\eta$-meson},  {\em Z. Phys.} {\bf C25} (1984)
  225--229.

\bibitem{Abele+1998}
{\bf Crystal Barrel} Collaboration, A.~Abele {et.~al.}, {\it Decay dynamics of
  the process $\eta \to 3 \pi^0$},  {\em Phys. Lett.} {\bf B417} (1998)
  193--196.

\bibitem{Tippens+2001}
{\bf Crystal Ball} Collaboration, W.~B. Tippens {et.~al.}, {\it Determination
  of the quadratic slope parameter in $\eta \to 3\pi^0$ decay},  {\em Phys.
  Rev. Lett.} {\bf 87} (2001) 192001.

\bibitem{Achasov+2001}
M.~N. Achasov {et.~al.}, {\it Dynamics of $\eta \to 3 \pi^0$ decay},  {\em JETP
  Lett.} {\bf 73} (2001) 451--452.

\bibitem{Bashkanov+2007}
M.~Bashkanov {et.~al.}, {\it Measurement of the slope parameter for the $\eta
  \to 3 \pi^0$ decay in the $pp \to pp \eta$ reaction},  {\em Phys. Rev.} {\bf
  C76} (2007) 048201, [\href{http://xxx.lanl.gov/abs/0708.2014}{{\tt
  arXiv:0708.2014}}].

\bibitem{Adolph+2009}
{\bf WASA-at-COSY} Collaboration, C.~Adolph {et.~al.}, {\it Measurement of the
  $\eta \to 3 \pi^0$ {D}alitz plot distribution with the {WASA} detector at
  {COSY}},  {\em Phys. Lett.} {\bf B677} (2009) 24--29,
  [\href{http://xxx.lanl.gov/abs/0811.2763}{{\tt arXiv:0811.2763}}].

\bibitem{Unverzagt+2009}
{\bf Crystal Ball at MAMI} Collaboration, M.~Unverzagt {et.~al.}, {\it
  Determination of the {D}alitz plot parameter $\alpha$ for the decay $\eta \to
  3 \pi^0$ with the {C}rystal {B}all at {MAMI-B}},  {\em Eur. Phys. J.} {\bf
  A39} (2009) 169--177, [\href{http://xxx.lanl.gov/abs/0812.3324}{{\tt
  arXiv:0812.3324}}].

\bibitem{Prakhov+2009}
{\bf Crystal Ball at MAMI} Collaboration, S.~Prakhov {et.~al.}, {\it
  Measurement of the slope parameter $\alpha$ for the $\eta\to 3\pi^0$ decay
  with the {C}rystal {B}all at {MAMI-C}},  {\em Phys. Rev.} {\bf C79} (2009)
  035204, [\href{http://xxx.lanl.gov/abs/0812.1999}{{\tt arXiv:0812.1999}}].

\bibitem{Ambrosino+2010}
{\bf KLOE} Collaboration, F.~Ambrosino {et.~al.}, {\it Measurement of the
  $\eta\to 3\pi^{0}$ slope parameter $\alpha$ with the {KLOE} detector},  {\em
  Phys.Lett.} {\bf B694} (2010) 16--21,
  [\href{http://xxx.lanl.gov/abs/1004.1319}{{\tt arXiv:1004.1319}}].

\bibitem{Schneider+2011}
S.~P. Schneider, B.~Kubis, and C.~Ditsche, {\it Rescattering effects in $\eta
  \to 3\pi$ decays},  {\em JHEP} {\bf 02} (2011) 028,
  [\href{http://xxx.lanl.gov/abs/1010.3946}{{\tt arXiv:1010.3946}}].

\bibitem{Colangelo+2006a}
G.~Colangelo, J.~Gasser, B.~Kubis, and A.~Rusetsky, {\it Cusps in ${K} \to 3
  \pi$ decays},  {\em Phys.Lett.} {\bf B638} (2006) 187--194,
  [\href{http://xxx.lanl.gov/abs/hep-ph/0604084}{{\tt hep-ph/0604084}}].

\bibitem{Bissegger+2008}
M.~Bissegger, A.~Fuhrer, J.~Gasser, B.~Kubis, and A.~Rusetsky, {\it Cusps in
  {$K_L \to 3 \pi$} decays},  {\em Phys.Lett.} {\bf B659} (2008) 576--584,
  [\href{http://xxx.lanl.gov/abs/0710.4456}{{\tt arXiv:0710.4456}}].

\bibitem{Bissegger+2009}
M.~Bissegger, A.~Fuhrer, J.~Gasser, B.~Kubis, and A.~Rusetsky, {\it Radiative
  corrections in {$K \to 3 \pi$} decays},  {\em Nucl.Phys.} {\bf B806} (2009)
  178--223, [\href{http://xxx.lanl.gov/abs/0807.0515}{{\tt arXiv:0807.0515}}].

\bibitem{Ananthanarayan+2001}
B.~Ananthanarayan, G.~Colangelo, J.~Gasser, and H.~Leutwyler, {\it Roy equation
  analysis of $\pi \pi$ scattering},  {\em Phys. Rept.} {\bf 353} (2001)
  207--279, [\href{http://xxx.lanl.gov/abs/hep-ph/0005297}{{\tt
  hep-ph/0005297}}].

\bibitem{Colangelo+2001}
G.~Colangelo, J.~Gasser, and H.~Leutwyler, {\it $\pi \pi$ scattering},  {\em
  Nucl.Phys.} {\bf B603} (2001) 125--179,
  [\href{http://xxx.lanl.gov/abs/hep-ph/0103088}{{\tt hep-ph/0103088}}].

\bibitem{Kaminski+2008}
R.~Kaminski, J.~Pelaez, and F.~Yndurain, {\it Pion-pion scattering amplitude.
  {III.} improving the analysis with forward dispersion relations and {R}oy
  equations},  {\em Phys.Rev.} {\bf D77} (2008) 054015,
  [\href{http://xxx.lanl.gov/abs/0710.1150}{{\tt arXiv:0710.1150}}].

\bibitem{Kampf+2011}
K.~Kampf, M.~Knecht, J.~Novotn\'y, and M.~Zdr\'ahal, {\it Analytical dispersive
  construction of $\eta\to3\pi$ amplitude: first order in isospin breaking},
  {\em Phys.Rev.} {\bf D84} (2011) 114015,
  [\href{http://xxx.lanl.gov/abs/1103.0982}{{\tt arXiv:1103.0982}}].

\bibitem{Colangelo+2009}
G.~Colangelo, S.~Lanz, and E.~Passemar, {\it A new dispersive analysis of $\eta
  \to 3 \pi$},  {\em PoS} {\bf CD09} (2009) 047,
  [\href{http://xxx.lanl.gov/abs/0910.0765}{{\tt arXiv:0910.0765}}].

\bibitem{Colangelo+2011a}
G.~Colangelo, S.~Lanz, E.~Passemar, and H.~Leutwyler, {\it Determination of the
  light quark masses from $\eta \to 3 \pi$},  {\em PoS} {\bf EPS-HEP2011}
  (2011) 304.

\bibitem{Lanz2011}
S.~Lanz, {\it Determination of the quark mass ratio {$Q$} from $\eta \to 3
  \pi$}.
\newblock PhD thesis, University of Bern, 2011.

\bibitem{Anisovich+1996}
A.~Anisovich and H.~Leutwyler, {\it Dispersive analysis of the decay $\eta \to
  3 \pi$},  {\em Phys.Lett.} {\bf B375} (1996) 335--342,
  [\href{http://xxx.lanl.gov/abs/hep-ph/9601237}{{\tt hep-ph/9601237}}].

\bibitem{Walker1998}
M.~Walker, {\it $\eta \to 3 \pi$},  Master's thesis, University of Bern, 1998.

\bibitem{Stern+1993}
J.~Stern, H.~Sazdjian, and N.~H. Fuchs, {\it What $\pi$-$\pi$ scattering tells
  us about chiral perturbation theory},  {\em Phys. Rev.} {\bf D47} (1993)
  3814--3838, [\href{http://xxx.lanl.gov/abs/hep-ph/9301244}{{\tt
  hep-ph/9301244}}].

\bibitem{Omnes1958}
R.~Omn\`es, {\it On the solution of certain singular integral equations of
  quantum field theory},  {\em Nuovo Cim.} {\bf 8} (1958) 316--326.

\bibitem{Bijnens+2002}
J.~Bijnens and J.~Gasser, {\it Eta decays at and beyond $p^4$ in chiral
  perturbation theory},  {\em Phys.Scripta} {\bf T99} (2002) 34--44,
  [\href{http://xxx.lanl.gov/abs/hep-ph/0202242}{{\tt hep-ph/0202242}}].

\bibitem{Kambor+1996}
J.~Kambor, C.~Wiesendanger, and D.~Wyler, {\it Final state interactions and
  {K}huri-{T}reiman equations in $\eta\to 3\pi$ decays},  {\em Nucl. Phys.}
  {\bf B465} (1996) 215--266,
  [\href{http://xxx.lanl.gov/abs/hep-ph/9509374}{{\tt hep-ph/9509374}}].

\bibitem{Gasser+1985}
J.~Gasser and H.~Leutwyler, {\it Chiral perturbation theory: expansions in the
  mass of the strange quark},  {\em Nucl. Phys.} {\bf B250} (1985) 465.

\bibitem{Ananthanarayan+2004}
B.~Ananthanarayan and B.~Moussallam, {\it Four-point correlator constraints on
  electromagnetic chiral parameters and resonance effective {L}agrangians},
  {\em JHEP} {\bf 06} (2004) 047,
  [\href{http://xxx.lanl.gov/abs/hep-ph/0405206}{{\tt hep-ph/0405206}}].

\bibitem{Kastner+2008}
A.~Kastner and H.~Neufeld, {\it The {$K_{\ell 3}$} scalar form factors in the
  standard model},  {\em Eur. Phys. J.} {\bf C57} (2008) 541--556,
  [\href{http://xxx.lanl.gov/abs/0805.2222}{{\tt arXiv:0805.2222}}].

\end{thebibliography}

\providecommand{\href}[2]{#2}\begingroup\raggedright\endgroup

\end{document}